\def\etal{{et al.\thinspace}}

\def\spose#1{\hbox to 0pt{#1\hss}}
\def\approxlt{\mathrel{\spose{\lower 3pt\hbox{$\sim$}}
        \raise 2.0pt\hbox{$$<$$}}}
\def\approxgt{\mathrel{\spose{\lower 3pt\hbox{$\sim$}}
        \raise 2.0pt\hbox{$>$}}}

\def\multleft#1{\hbox to size{\vbox {\halign {\lft{##}\cr #1}}\hfill}\par}
\def\multright#1{\hbox to size{\vbox {\halign {\rt{##}\cr #1}}\hfill}\par}

\def\today{\ifcase\month\or January\or February\or March\or April\or May\or
      June\or July\or August\or September\or October\or November\or December\fi
      \space\number\day, \number\year}
\def\s{\hbox{\phantom{5}}}      %one space
\def\boxit#1{\vbox{\hrule\hbox{\vrule\kern3pt\vbox{\kern3pt
          #1 \kern3pt}\kern3pt\vrule}\hrule}}

\def\cm{{\rm\thinspace cm}}

\def\g{{\rm\thinspace g}}
\def\ga{{\rm\thinspace gauss}}

\def\km{{\rm\thinspace km}}

\def\Lsun{\hbox{$\rm\thinspace L_{\odot}$}}

\def\Mpc{{\rm\thinspace Mpc}}
\def\Msun{\hbox{$\rm\thinspace M_{\odot}$}}

\def\s{{\rm\thinspace s}}
\def\yr{{\rm\thinspace yr}}

\def\kmps{\hbox{$\km\s^{-1}\,$}}

\def\Msunpyr{\hbox{$\Msun\yr^{-1}\,$}}

\def\mic{\hbox{\rm\thinspace $\mu$m}}
\def\mJy{{\rm\thinspace mJy}}
\def\ng5{12}

\documentstyle[epsfig,psfig]{mn}
\begin{document}
\hsize=6truein

\title[Submm observations of $z>5$ quasars]
{Quasars as probes of the submillimetre cosmos at {$\mathbf z>5$}:\\
I. Preliminary SCUBA photometry}
\author[R.S. Priddey, K.G. Isaak, R.G. McMahon et al.]
{\parbox[]{6.5in}
{Robert S. Priddey$^1$,%\footnote{email: r.priddey@ic.ac.uk}, 
Kate G. Isaak$^2$, Richard G. McMahon$^3$,
E.I. Robson$^4$, C.P. Pearson$^{1,5}$}
\\
{\it 
$^1$ Imperial College, Blackett Laboratory, Prince Consort Road, 
London SW7 2BZ}\\
{\it
$^2$ Cavendish Laboratory, Madingley Road, Cambridge CB3 UK}\\
{\it 
$^3$ Institute of Astronomy, Madingley Road, Cambridge CB3 0HA, UK}\\
{\it
$^4$ UK Astronomy Technology Centre, Blackford Hill, Edinburgh EH9 3HJ, UK}\\
{\it 
$^5$ School of Physical Sciences, University of Kent at Canterbury, 
Canterbury, UK}
}

%%%Accepted August 5 2003; submitted June 27 2003; in original form April 4

\date{To appear in MNRAS Letters.}

\maketitle

\begin{abstract}
We present submillimetre (submm) continuum observations
of a sample of some of the highest redshift quasars currently known,
made with the SCUBA bolometer array on the JCMT.
\ng5\ of the sample have redshifts greater than five, and 
two have $z\ge6$; the median redshift of the sample is 5.3.
Two of the $z>5$ objects are strong (6$\sigma$) detections, and are 
bright sources with $S_{850\mic}>10$\mJy. Another firm (5$\sigma$) 
detection is obtained
for the $z=5.7$ quasar SDSS J1044$-$0125; and SDSS J1306+0356, at $z=6.0$,
is detected with a signal-to-noise ratio $\approx$4.
We have obtained sensitive ($\sigma\approx$1.5mJy) 
upper limits 
for much of the remainder of the sample, including the $z=6.3$ quasar 
SDSS J1030+0524.

Submm spectral indices measured for two of the sources ($\alpha\approx3.3$)
are consistent with
thermal reradiation from dust, rather than from synchrotron emission.
Sensitive upper limits at 450$\mu$m imply that the dust is cool, requiring
large dust masses ($10^{8-9}\Msun$) to account for the observed fluxes---
suggesting substantial   
prior star formation, even at $z=6$ when the Universe was only 1.0Gyr old.

\end{abstract}


\section{INTRODUCTION}
Forty years ago, an accurate optical identification was obtained for
the curious radio source 3C273 (Hazard, Mackey \& Shimmins, 1963).
Subsequent spectroscopy revealed an unexpectedly high redshift, $z=0.16$,
heralding a dramatic new type of astrophysical phenomenon--- 
dubbed the ``quasi-stellar object'' (QSO).
Ever since, 
the extremely luminous, compact emission from quasars 
has
rendered them almost unrivalled as beacons to the distant Universe---
whether as spotlights shining through absorption systems, pinpoints 
fixing their
host galaxies or as tracers of large scale structure.

Recently, however, dedicated searches for {\it galaxies} at the highest
redshifts have unveiled new populations of star-forming sources
at $z>5$ and $z>6$ (e.g. Hu et al. 2002; Stanway, Bunker \& McMahon 2003).
Nevertheless, the latest AGN (Active Galactic Nuclei)
surveys keep pace, and efficiently yield 
quasars at $z>5$ (refs. under Fan et al.; Sharp et al. 2001),
with the promise of many more to be discovered 
over the coming years.
That AGN remain important cosmological tools 
was recently highlighted by
the first evidence for Gunn--Peterson absorption, discovered 
in the spectrum of a $z=6.3$ quasar (Becker et al. 2001).
The high-redshift quasars of this new generation 
stand tantalizingly as signposts
to the epoch of reionization.

The very existence of luminous quasars at $z>5$, indeed, 
poses a challenge to
structure formation theory. 
Their supermassive black hole engines
($M\sim10^9\Msun$), and the fuel reservoir required to supply them
($\dot{M}\sim100\Msunpyr$), constitute a large mass concentration
already in place only a gigayear after recombination.
The Cold Dark Matter cosmogony, in contrast,
predicts that the first objects to form were of 
predominantly low mass ($\sim10^6\Msun$).
High-redshift quasars therefore offer a 
a test of hierarchical cosmology
operating at its extreme (Efstathiou \& Rees, 1988; Turner, 1991).
The signatures of high overdensity,
for example the biased evolution of the surrounding region, 
can in principle be detected observationally
by targetting quasars and their environs
(e.g. in the submm, Ivison et al. 2000).

\medskip\noindent
Here, we report continuum submillimetre observations, made with the
SCUBA bolometer array on the James Clerk Maxwell Telescope, of a sample 
of some of the highest-redshift quasars currently known.
These observations form part of a larger project to establish the 
fiducial properties of cool dust emission from
radio-quiet, optically-selected quasars 
over a wide range in redshift ($z<1$: Isaak et al., in prep.; 
$z=2$: Priddey et al. 2003a; $z>4$: McMahon et al. 1999, Isaak et al. 2002). 
Our earlier work suggested that a significant fraction of optically-bright
quasars are ultra- or even hyper-luminous at far-infrared wavelengths.
Additionally, 
no significant difference between the $z=2$ and $z>4$ samples was detected,
contrasting with the SCUBA radio galaxy survey by Archibald et al. (2001).
Several lines of evidence suggest that the submm light is powered by
star formation, and multiband follow-up is in progress to test this
hypothesis.
Observation beyond $z>5$, 
reaching yet further back in cosmic time,
is a vital extension to the project.
Are conditions at the highest redshifts favourable, or adverse, to
the formation of luminous, dusty sources? 
Do we catch the quasars during an even more turbulent phase of 
youth (gas-rich, 
violently star-forming), or does there exist a redshift cut-off beyond which
too little dust has formed for the object to be a SCUBA source?

This Letter is the first in a series of papers discussing 
submm sources at $z>5$.
Here we present initial submillimetre photometry of 
the sample, leaving a presentation of follow-up data, and a more thorough
account of interpretation,
to forthcoming works (Isaak et al. in prep.; Robson et al. in prep.).
Throughout, we assume a flat, $\Lambda$-dominated cosmology
$\Omega_M$=0.3, $\Omega_{\Lambda}$=0.7 with
$H_0$=65\kmps\Mpc$^{-1}$. Additional quantities (in parentheses)
are for an Einstein--de Sitter (EdS) universe with $H_0$=50\kmps\Mpc$^{-1}$.

\section{OBSERVATIONS}
The observed quasars are listed in Table 1.

\subsection{The sample}
The parent sample consisted of all quasars above a redshift
of 4.90 that were known at the time of observation (i.e. late 2001). 
The targets were
drawn from the Sloan Digital Sky Survey (SDSS: see Fan et al. refs.), the
Isaac Newton Wide Field Survey (WFS: Sharp et al., 2001) and the single
object (RD J0301+0020) reported by Stern et al. (2000).
Targets were prioritized solely on the basis of redshift, with target
visibility providing the only additional constraint.

Absolute $B$ magnitudes are calculated from
1450\AA\ continuum fluxes obtained from published spectra, 
assuming an optical spectral
index $\alpha=-0.5$. In Table 1, the primary values are for the $\Lambda$
cosmology, and those for $\Omega_M=1$ are given in parentheses.
In our previous submm quasar surveys (McMahon et al., 1999;
Isaak et al., 2002; Priddey et al., 2003), we had
targetted quasars
brighter than $M_B=-27.5$ (based on the EdS cosmology). 
At $z>5$, too few sources are known for this stringent criterion
to be maintained:
the median and mean magnitude are each now $M_B=-26.6$ (again EdS).

\subsection{JCMT--SCUBA submillimetre data}
Data for SDSS J1044$-$0125 were obtained through service observing 
time in July 2000 and January 2001.
The bulk of the rest of the data were obtained during the period
autumn 2001--summer 2002, 
through a combination of "flexible" scheduling, 
the scheduled time itself and Director's Discretionary Time. 

SCUBA was employed in photometry mode, with the wide 850:450 filter set, 
and a standard 60 arcsec chop in azimuth at 7.8Hz. The source was placed on the
central bolometer (H7, C14), and the median of the remaining 
(quiet) bolometers 
was used for additional sky removal. 
Flux calibration was performed against the planets
Mars and Uranus, and against the standard continuum calibrators CRL618,
OH231.8, IRC10216 and 16293$-$2422. 
Telescope pointing was checked frequently. Sky opacity was monitored
via regular skydips at 850 and 450$\mu$m, and continuously at 225GHz via the 
JCMT Water Vapour Monitor and the CSO Tau Meter:
over the whole set of observations, $\tau_{225}$ ranged between
$0.04<\tau_{225}<0.14$.
Data were reduced with both the ORAC-DR pipeline and the SURF software.

Both SDSS J1044$-$0125 and SDSS J1306+0356 were reobserved,
after showing some inconsistencies in their initial
datasets: the detections were confirmed in each case.
For example, 
during initial observation of SDSS J1306+0356, in November 2001, 
the final two blocks of integrations
were inconsistent with the data taken over the preceding few hours. 
Their mean signals were zero or negative,
where the others had throughout been strongly positive.
This marked change in consistency coincided not only with deteriorating weather
(seeing poorer than 4 arcsec), but
with sunrise, and consequent fleeting changes in atmosphere, 
dish figure and absolute pointing accuracy.
We reobserved the quasar under good, stable conditions
(seeing 0.5 arcsec, $\tau_{\rm 225GHz}=0.065$), 
conditions in March 2002.
The signal thereby obtained was consistent with the first seven blocks 
from the previous November.
The final flux quoted in Table 1 therefore consists of the 
November data minus the last two blocks, plus the March data.
Adding back the suspect data would give $S_{\rm 850\mu m}$=3.1$\pm$0.9mJy--- 
a signal-to-noise which is still greater than three.

SCUBA consists of two detector arrays, dedicated respectively
to long- and short-wavelength observations--- in this case 850 and 450$\mu$m.
The atmospheric transmission decreases with frequency,
hence the long-wave array is the usual primary instrument.
Nevertheless, 450$\mu$m data are obtained {\it gratis}, and they are here  
of particular interest, for they provide important constraints on the
physical temperature of the dust.
450$\mu$m fluxes are thus included in Table 1 for reference, however
none of the sources was detected in the band.
Observations were made under a wide range of atmospheric
conditions (450$\mu$m opacities $0.75<\tau_{\rm450\mu m}<3.0$),
resulting in a very heterogeneous set of RMS values, between 
10 and 100mJy.
We emphasise that accurate calibration of these short-wavelength
observations is possible only under the very best observing conditions.
For comparison, the relative errors in flux calibration with the
850 and 450 arrays are typically 5--10 percent and $\sim20$ percent,
respectively.

\section{RESULTS}

\begin{table*}
\caption{Quasars at $z>5$ observed with SCUBA}
%\begin{center}
\begin{tabular}{lcccccrrl}
Target name & $z$ & $M_B$ & RA & Dec & Number of & $S_{850\mic}$ & 
$S_{450\mic}$ & Ref. \\
& & & (J2000) & (J2000) & integrations& (mJy) & (mJy) & \\
(1) & (2) & (3) & (4) & (5) & (6) & (7) & (8) & (9)\\
\hline
SDSS J1030+0524 & 6.28 & $-$27.8 ($-$27.4) & 10 30 27.10 & +05 24 55.0 & 
475 & 1.3$\pm$1.0&$-21\pm$10 & 1\\
SDSS J1306+0356 & 5.99 & $-$27.8 ($-$27.4) & 13 06 08.26 & +03 56 26.3 & 
490 & {\bf 3.7$\pm$1.0}&$-7\pm$14 & 1\\
SDSS J0836+0054 & 5.82 & $-$28.5 ($-$28.1) & 08 36 43.85 & +00 54 53.3 & 
250 & 1.7$\pm$1.5 & $-24\pm$10 & 1\\
SDSS J1044$-$0125 & 5.74 & $-$28.1 ($-$27.7) & 10 44 33.04& $-$01 25 02.2& 
500 & {\bf 6.1$\pm$1.2} & n/a$^{a}$ & 2\\
RD J0301+0020 & 5.50 & $-$23.1 ($-$22.7) & 03 01 17.01 & +00 20 26.0 &  
200 & 1.9$\pm$1.5& $-5\pm$18& 3\\
SDSS J0231$-$0728 & 5.41 & $-$28.0 ($-$27.6)& 02 31 37.65 & $-$07 28 54.5 & 
200 & 1.8$\pm$1.6& 2$\pm$15 & 4\\
SDSS J1208+0010   & 5.27 &  $-$26.7 ($-$26.3) & 12 08 23.82 & +00 10 27.7 & 
200 & $-2.0\pm$2.5 & $-64\pm$102 & 5\\
WF J2245+0024   & 5.17 & $-$25.3 ($-$24.9)& 22 45 24.28 & +00 24 14.6 &
180 & 2.3$\pm$1.6 & 2$\pm$15 & 6\\
SDSS J0913+5919$^{b}$ & 5.11 & $-$26.3 ($-$26.0)& 09 13 16.56 & +59 19 21.5 &
200 & 2.8$\pm$1.8 & 4$\pm$28 & 4\\
SDSS J1204$-0021$  & 5.11 & $-$28.0 ($-$27.6)  & 12 04 41.70 & $-$00 21 49.6 & 
200 & 4.2$\pm$2.0 & $-104\pm$50 & 7\\
SDSS J0756+4104 & 5.09 & $-$27.0 ($-$26.6)& 07 56 18.14 & +41 04 08.6 & 
100 & {\bf 13.4$\pm$2.1}& 14$\pm$19 & 4\\ 
SDSS J0338+0021 & 5.07 & $-$27.0 ($-$26.7)& 03 38 29.31 & +00 21 56.3 & 
100 & {\bf 11.9$\pm$2.0}& 5$\pm$16 & 8\\
SDSS J2216+0013 & 4.99 & $-$26.9 ($-$26.6)& 22 16 44.02 & +00 13 48.3 &
250 & 1.7$\pm$1.4 & 14$\pm$19 & 4\\
WF J1612+5255 & 4.95 & $-$26.4 ($-$26.1) & 16 12 53.10 & +52 55 43.5 & 
250 & 2.1$\pm$1.9 & 21$\pm$50 & 6\\
\hline

\end{tabular}
%\end{center}
\begin{flushleft}
{\bf Refs:} 1. Fan et al. 2001; 2. Fan et al. 2000b; 
3. Stern et al. 2000; 4. Anderson et al. 2001; 5. Zheng et al. 2000;\\
6. Sharp et al. 2001; 7. Fan et al. 2000a; 8. Fan et al. 1999
\\
\smallskip
$^{a}$ Observations made during high atmospheric opacity
(Weather Grade 3): short-wave data consequently very poor\\
$^{b}$ Radio source: $S_{\rm 1.4GHz}$=17.7mJy (NVSS catalogue)
\end{flushleft}
\end{table*}

Two of the targets, {\bf SDSS J0756+4104} and {\bf SDSS J0338+0021} 
(both $z=5.1$)
are strikingly bright submm sources, with $S_{\rm 850\mu m}>10$mJy. 
The latter was also detected at 1.2mm by Carilli et al. (2001),
$S_{1.2\rm mm}=3.7$mJy. The 850$\mu$m flux is consistent with a thermal
spectrum (see Figure 1 and Section 4.1).
{\bf SDSS J1044$-$0125} ($z=5.7$) is a moderately bright detection,
$S_{850}=7$mJy.
Its 1.2mm flux of around 2mJy 
(Isaak et al., in prep.) is consistent with the steep
Rayleigh--Jeans tail of a thermal greybody spectrum from dust.
This source and its environs have 
formed the target of a range of follow-up observation, reported in
a companion paper (Isaak et al., in prep.).
It is notable that SDSS1044 is a broad absorption line (BAL) quasar,
a fact drawn upon to account for its X-ray weakness, relative to the optical,
as measured by Brandt et al. (2001).
It is conceivable that the gaseous outflow responsible for the 
optical absorption also absorbs the X-rays; and that dust 
embedded within the gas gives rise to the submm emission.
{\bf SDSS J1306+0536} ($z=6.0$) was detected at 850$\mu$m 
with a significance of 3.7$\sigma$ (see also Section 2.2).

\medskip\noindent
RD J0301+0020 (z=5.50) was detected at 1.2mm in a very deep IRAM--MAMBO 
observation by 
Bertoldi \& Cox (2002): the 850$\mu$m limit presented here
is consistent with a thermal spectrum given
their $S_{1.2}=0.87\pm0.20$mJy. RD J0301+0020
stands out from the other $z>5$ quasars by virtue of its extremely
low optical luminosity, $M_B=-23.1$. Using the Elvis et al. (1994)
bolometric correction from $\nu L_B$ ($C_B=12$) implies 
$L_{bol}\approx2\times10^{12}\Lsun$.
This is comparable to the far-infrared luminosity derived from
the millimetric flux, if we assume a canonical cool dust spectrum
($T=40-50$K). 
Assuming that the quasar (and not some foreground object) 
is the MAMBO source, then either 
the AGN is absorbed, and we are seeing reprocessed emission
in the far-infrared, or the $L_{\rm FIR}$ derives from star formation
(or, indeed, a combination of the two).

\subsection{Caveat emptor: the effects of lensing}
In neither Table 1 nor Table 2 do we attempt to correct the observed
fluxes, and their derived quantities, for a gravitational 
lensing magnification.
At $z>5$, one would expect the optical depth to lensing to be significant;
Wyithe \& Loeb (2002) additionally point out that the $z>5$ SDSS survey
selects quasars on the steep, bright tail of the luminosity function,
rendering it prone to magnification bias.
Gravitational lensing is therefore a serious problem whose effects must
be addressed on a source-by-source basis. 
For now we can only do so statistically: for our two $z>5.5$ detections,
SDSS J1306 and SDSS J1044, Wyithe \& Loeb (2002)
estimate a 7--30 percent probability of multiple imaging,
with median (mean) magnifications between 1.1 (5) and 1.2 (30).

Schwartz (2002) examined the {\it Chandra}--ACIS images of the three 
highest-redshift objects in this sample. The six photons comprising the
SDSS J1030 detection are widely distributed, statistically inconsistent 
with a point source. This is a possible signature of lensing, however
the counts are too few to differentiate between a single extended source
or multiple point sources.
In contrast, Fan et al. (2003) report that {\it HST}--ACS 
images of SDSS J1030 and J0836 are consistent with 
unresolved point sources. Their ground-based imaging of SDSS J1306 
promotes a similar conclusion.

\begin{table*}
\caption{Derived properties of the sample 
(detections and deep ($\sigma\le1.5$mJy) limits). Numbers in parentheses
are for the EdS cosmology (except for $\dot{M}_*$(min), where the
numbers for each cosmology are approximately equal). RD0301 is counted among
the deep 850$\mu$m non-detections, but we have used its 1.2mm flux to 
calculate the parameters.
N.B. $L_{\rm FIR}$ and $\dot{M}_*$ are {\it not} independent.}
\begin{tabular}{lcclllll}
Source & $z$ & $t(\infty)-t(z)$ & $M_d$ & $\dot{M}_*$(min) & $L_{\rm FIR}$
& $M_{bh}$ & $\dot{M}_{acc}$ \\
       &     & Gyr & $10^8$\Msun & $\Msunpyr$ & $10^{10}\Lsun$
& $10^9\Msun$  & $\Msunpyr$ \\ 
\\
\hline

SDSS J1306+0356   & 5.99 & 0.99 (0.70) &2.6 (1.8)  &26 &520 (370) 
& 4.4 (3.0) & 95 (65) \\
SDSS J1044$-$0125 & 5.74 & 1.04 (0.74) &4.2 (3.0)  &41 &870 (610) 
& 5.6 (4.0) & 125 (85) \\
SDSS J0756+4104   & 5.09 & 1.21 (0.86) &9.6 (6.9)&80&1970 (1410) 
& 2.1 (1.4) & 45 (30) \\
SDSS J0338+0021   & 5.07 & 1.22 (0.87) &8.5 (6.1) &70&1750 (1250) 
& 2.1 (1.4) & 45 (30) \\
%\\
SDSS J1030+0524   & 6.28 & 0.93 (0.66) &$<$1.4 (1.0)&---&$<$280 (200) 
& 4.4 (3.0) & 95 (65) \\
SDSS J0836+0054   & 5.82 & 1.02 (0.73) &$<$2.1 (1.5)&---&$<$430 (300) 
& 7.6 (5.2) & 165 (115) \\
%RD J0301+0020     & 5.50 & 1.10 (0.78) &$<$3.4 (2.4)&---&$<$430 (310)\\
RD J0301+0020     & 5.50 & 1.10 (0.78) & 1.4 (1.0) & 13 & 290 (200) 
& 0.06 (0.04) & 1.4 (0.9) \\
SDSS J2216+0013   & 4.99 & 1.25 (0.89) &$<$2.0 (1.4)&---&$<$410 (300) 
& 1.9 (1.4) & 40 (30) \\

\end{tabular}
\end{table*}

\begin{figure}
\begin{center}
\epsfig{figure=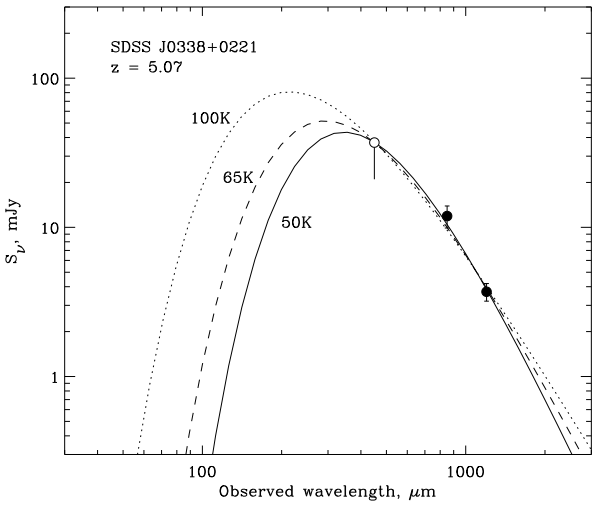,width=100mm}
\end{center}
\caption{Constraining the dust temperature of SDSS0338. The 1.2mm point is
from Carilli et al. (2001), the 850 and 450$\mu$m from this work.
Curves are for the maximum dust temperatures (labelled) consistent 
with the 450$\mu$m upper limit (plotted as signal plus 2-$\sigma$ with a 
$\sigma$-length bar), for three greybody indices: $\beta=2.0$ (solid), 
$\beta=1.5$ (dashed) and $\beta=1.0$ (dotted). 
The steep submm--mm slope renders a better fit for a high $\beta$, 
thus favouring a low temperature.}
\end{figure}

\section{DISCUSSION}

\subsection{Dust, metals and star formation at $z>5$}
The mm--submm spectral indices measured for SDSS J1044$-$0125
and J0338+0021--- each $\alpha^{mm}_{submm}\approx3.3$---
are consistent with
the Rayleigh--Jeans tail of thermal emission from cool dust.
Such a spectral energy distribution (SED) would be 
expected to peak at a rest wavelength 
$\approx50-100\mu$m; similar, for example, to star-forming 
galaxies such as M82.
Indeed, existing 450 and 350$\mu$m detections of $z\approx4$ 
quasars suggest that
the SED tends to a plateau (see e.g. Priddey \& McMahon, 2001 (PM01)) at 
around 100$\mu$m rest-frame: all but the brightest
sources would lie below the short-wavelength limit that we could detect.
At $z>5$, then, a 450$\mu$m {\it detection} would imply that the dust is hot;
conversely, the deep limits obtained for the bright 850$\mu$m detections
are valuable constraints on their dust temperature.
Again considering J0338+0021, 450 upper limit requires $T<100$K for
a greybody index $\beta=1$, and $T<50$K for $\beta=2$, the higher $\beta$
favoured by the steep mm--submm slope (Figure 1). 
(However we warn of the potentially considerable systematic uncertainties
in the 450$\mu$m calibration.)
In the following, we shall therefore assume 
the SED of PM01, namely $\beta$=2, $T$=40K.

Temperature uncertainty notwithstanding,
determination of 
the dust opacity ($\kappa$) has, in the past, presented additional
uncertainties to a derivation of a dust mass from a submm flux. 
We adopt a value $\kappa(125\mu\rm m)$=30$\cm^2\g^{-1}$, at a normalisation
wavelength (125$\mu$m) corresponding to observed 850$\mu$m at $z$=5.8.
This is consistent with the $\kappa(125)$ determined by Hildebrand (1983).
Extrapolated to longer wavelengths assuming $\kappa\propto\lambda^{-2}$,
it is also consistent with a Galactic mixture of Draine \& Lee (1984) grains
as well as with more recent measurements,
e.g. the 850$\mu$m value of James et al. (2002).
For our $z>5$ sample, it yields masses in the range $10^{8-9}\Msun$ (Table 2).

The youth of the universe--- 1Gyr at $z=5$--- imposes a constraint on the 
mechanisms of dust production, for it is comparable to the
evolutionary timescale of the stars in whose atmospheres the dust is
believed to condense. An alternate method of manufacture, 
on the winds of Type II supernovae, requires much shorter timescales,
but its efficacy has yet to be proven.
Crudely assuming an overall dust enrichment rate per mass forming stars of 
around 1 percent,
and an absolute upper limit of 1Gyr available for star formation, 
then a dust mass of $10^{8-9}\Msun$ would
require a minimum average star formation rate of $\sim100\Msunpyr$. It is thus
plausible that star-forming activity is responsible for some or all 
of the observed submm luminosity. 
In Table 2, the quantity $\dot{M}_*$(min) is the minimum star formation rate
necessary to produce the estimated dust masses, averaged 
the lifetime of the universe. Alternatively, $L_{\rm FIR}$
could be used to estimate the 
{\it instantaneous} star formation rate, 
$\frac{\dot{M}_*}{\Msunpyr} = \Psi\frac{L_{\rm FIR}}{10^{10}\Lsun}$,
where $\Psi$ is a constant of order unity, depending on the stellar 
initial mass function (IMF).
This quantity represents an upper limit on the star formation rate,
due to the potential of AGN contribution to dust heating.
Nevertheless it is plausible that the dust we observe was formed
during production of a substantial fraction of stars in the quasar's
host galaxy.

Morgan \& Edmunds (2003) present a detailed model of dust synthesis and
discuss its implications for very high redshift submm sources such
as the current sample. Their model implies that, in the absence of 
supernovae as a source of dust, the star formation efficiency must
be very high to yield, by $z=5$, the masses we observe.

\subsection{The growth of supermassive black holes}
The four highest-redshift quasars in the sample are all very 
luminous in the optical, $M_B\approx-28$.
The assumption that this luminosity is supplied by Eddington-limited accretion
then requires a black hole of at least 5$\times10^9\Msun$ (Table 2).
How does a black hole of such mass form? Imposing the maximum 
available time of $\sim$1Gyr implies an average accretion rate of
5\Msunpyr. 
In comparison, 
assuming an (optimistic) radiative efficiency $\epsilon=0.1$
(where $L_{\rm bol}=\epsilon \dot{M} c^2$), 
an (instantaneous) accretion rate of $\ga100\Msunpyr$ is required to fuel 
the bolometric power inferred from the optical.
The $e$-folding time for Eddington accretion is $\approx0.5\epsilon$Gyr;
starting from a $10^{3-6}\Msun$ seed (e.g. Haehnelt, Natarajan \& Rees, 1998), 
an observed redshift of $z$=6 would
require a formation redshift $z$=10--18 in the $\Lambda$ cosmology,
but an upper limit longer than the age of the Universe in the matter-dominated
cosmology.

In the local universe, quiescent supermassive black holes and their
surrounding stellar bulges correlate in their mass and velocity
dispersion 
(Magorrian et al. 1998; Ferrarese \& Merrett 2000; Gebhardt et al. 2000). 
It is likely, therefore, that their
formation processes--- the one through accretion and an AGN phase,
the other perhaps through a dusty starburst--- are closely linked.
What can we conclude concerning the coevolution of a quasar and
its host galaxy, given these data?
Both PM01 and Archibald et al. (2002) consider this matter. They
propose a simple model in which a black hole exponentiating
according to the Eddington rate, and an elliptical host galaxy
forming stars and dust at roughly constant rate,
conspire to produce a submm-luminous AGN just before the gaseous
fuel supply is exhausted.
In a future paper,
we will consider a wider range of possibilities.

%%%%%%

\section{CONCLUSIONS AND FUTURE WORK}
We have presented SCUBA 850$\mu$m observations a sample of quasars at $z>5$---
where the youth of the Universe itself ($\approx$1Gyr) starts to become a
constraint on the histories of accretion and star formation---
detecting four sources at 850$\mu$m.
The few 1.2mm--850$\mu$m spectral indices that we can determine are
consistent with thermal reradiation from dust; and deep 450$\mu$m limits
imply that the dust is cool ($T<100$K), consistent with 
their siblings at $z=4$ (Priddey \& McMahon, 2001).
Although the sample is much smaller, and its median optical luminosity
fainter, 
its submm properties seem tentatively similar to those
of radio-quiet quasars we have studied at $z=4$ and $z=2$ 
(Isaak et al. 2002; Priddey et al. 2003a).

It is plausible that we are observing extreme, high-$\sigma$ peaks in the
overdensity distribution. The quasar host galaxies are likely massive, 
gas-rich and actively forming stars. We are pursuing a range of
follow-up observations designed to investigate such a scenario.
For example, submillimetre imaging of the fields of high-redshift AGN
has been used (e.g. Ivison et al. 2000) to test for biased galaxy formation.
In this spirit, we have obtained SCUBA jiggle maps of three
of the sources from the current sample, and our findings will be reported in 
a forthcoming paper (Isaak et al. in prep.).

An important quantifier of the evolutionary state of the sources themselves
will be obtained through detection of carbon monoxide emission lines,
which would reveal the existence of reservoirs of molecular gas
indicative of star formation.
Simultaneously,
it will be essential to improve the constraints on the 
infrared/submm SEDs of these quasars, to confirm the presence
of dust, differentiate between AGN- and starburst-powered
components and improve estimates of the dust mass and star formation
rate.
And of no less importance is the need to obtain high 
resolution optical/near-infrared
images of the quasars, to assess the probable effects of gravitational
lensing.

\medskip\noindent
It is an oft-cited consequence of the
negative submm $K$-correction that it hypothetically
permits an unidentified SCUBA source to lie at a redshift as high
as ten, without being intrinsically more luminous than a local
object like Arp220.
The difficulty of optically identifying such a source has 
prevented this claim becoming more than hypothetical.
Now, however, the submillimetre study of samples of quasars at $z>5$ provides 
a ready means of inferring the properties-- indeed the very existence-- 
of dust within one gigayear of the Big Bang.

\section{ACKNOWLEDGMENTS}
We are grateful to the observers
who gathered much of the data in ``flexible'' time, 
and to the TSSs and JAC staff (notably Jim Hoge, Ed Lundin and Iain Coulson) 
for providing efficient operation during our observing run.
We thank Matt Fox for his 
presence through the March 2002 run. 
We are indebted to 
the anonymous referee for making valuable comments.
The JCMT is operated by JAC, Hilo, on behalf 
of the parent organisations of the Particle Physics and Astronomy Research 
Council in the UK, the National Research Council of Canada and 
The Netherlands Organisation for Scientific Research.

\end{document}